**RESEARCH**

**Open Access**

# NBA: defensive distillation for backdoor removal via neural behavior alignment

Zonghao Ying[1,2] and Bin Wu[1,2*]

**Abstract**

Recently, deep neural networks have been shown to be vulnerable to backdoor attacks. A backdoor is inserted into neural networks via this attack paradigm, thus compromising the integrity of the network. As soon as an attacker presents a trigger during the testing phase, the backdoor in the model is activated, allowing the network to make specific wrong predictions. It is extremely important to defend against backdoor attacks since they are very stealthy and dangerous. In this paper, we propose a novel defense mechanism, Neural Behavioral Alignment (NBA), for backdoor removal. NBA optimizes the distillation process in terms of knowledge form and distillation samples to improve defense performance according to the characteristics of backdoor defense. NBA builds high-level representations of neural behavior within networks in order to facilitate the transfer of knowledge. Additionally, NBA crafts pseudo samples to induce student models exhibit backdoor neural behavior. By aligning the backdoor neural behavior from the student network with the benign neural behavior from the teacher network, NBA enables the proactive removal of backdoors. Extensive experiments show that NBA can effectively defend against six different backdoor attacks and outperform five state-of-the-art defenses.

## Introduction

Recent years have seen the use of deep learning for a wide range of critical tasks, such as autonomous vehicle driving (Grigorescu et al. 2020; Muhammad et al. 2021), facial recognition (Hu et'al. 2015; Wang and Guo 2021), machine translation (Costa-jussà 2018; Koehn 2020), etc. As deep learning expands its application scope, its security issues are also garnering increased attention (Berman et al. 2019; Liu et al. 2021; Guowen et al. 2019). Deep neural networks are regarded as key components of deep learning, and their security has always been emphasized in research. It is expensive and time consuming to train a deep neural network, so many users use Machine Learning as a Service (MLaaS) (Ribeiro et al. 2015) or directly download post-trained networks from the Internet. In this case, a third party handles the training of the network. An honest third party will train normally and return a clean model, however, there is also the possibility for a malicious third party to manipulate the training process and return a tainted model. Due to the black-box nature of neural network (Rudin 2019), users cannot determine whether the model has been maliciously modified. Service features of MLaaS and the black-box nature of the models offer the possibility of backdoor attack.

The backdoor attack (Gao et al. 2020) consists of two phases, namely the implanting phase and the activating phase. A backdoor is implanted during the training of the neural network, for example by tampering with the training data, and it is then activated during the testing of the network. Backdoor attacks have the main characteristic that the network will make specific incorrect predictions only when triggers are presented in the input, otherwise it behaves normally. As is shown in Fig. 1, the "STOP" sign will be predicted as "LIMIT 50" when the image recognition system predicts an image stamped with a trigger. When the system is applied to auto driving, this kind of backdoor behavior could contribute to serious traffic accidents. As mentioned earlier, a malicious third party is well positioned to implant a backdoor and return the backdoored network to the user. Users are also provided with a partially clean data set when they receive the network in order to test whether the it performs as expected. Nevertheless, the backdoor in the network cannot be activated by clean data, i.e., a user cannot determine whether the network contains a backdoor.

*Correspondence:
Bin Wu
wubin@iie.ac.cn
[1] State Key Laboratory of Information Security, Institute of Information Engineering, Chinese Academy of Sciences, Beijing, China
[2] School of Cyber Security, University of Chinese Academy of Sciences, Beijing, China

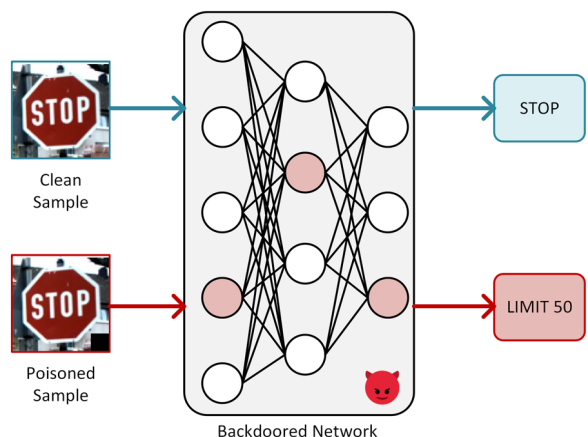

**Fig. 1** Example of a backdoor attack

The defense based on knowledge distillation is currently considered to be the most effective method for mitigating backdoor attacks. NAD [57] was the first to introduce knowledge distillation into backdoor defense. It utilizes attention features to represent the neural network's internal neuron activation information and achieves backdoor defense by aligning the intermediate layer attention features of the student network and the teacher network. The limitation of NAD is that it only involves same-order attention features during knowledge distillation, while the correlation among attention features of different orders is ignored. On this basis, ARGD[58] proposes the attention relation graph, which fully considers and utilizes the relationship between attention features of different orders. As a result, the defense performance is further improved. They have a common limitation, that is, they only optimize knowledge representation, and this knowledge representation is too single. Knowledge distillation was originally proposed because of the need to quantify the network, so we argue that simply optimizing knowledge representation is far from enough to defend against backdoors.

In this work, we propose a new defense mechanism called NBA, which simultaneously optimizes knowledge representation and training samples according to the characteristics of backdoor defense. In terms of knowledge representation, NBA defines and extracts three types of neural behaviors from within the neural network to fully represent the knowledge of the network. By optimizing the corresponding loss function, the student network can be encouraged to align its neural behavior with that of the teacher network, resulting in better training results. In contrast, the knowledge representation used by NAD and ARDG can essentially be regarded as one kind of neural behavior used by the NBA. In terms of training samples, we construct pseudo poisoned samples and input them to the student network. After the backdoor neural behavior is exposed, NBA can remove the backdoor more thoroughly. Based on the above optimizations, NBA can achieve better defensive performance than NAD and ARGD.

In summary, we make the following contributions:

- We propose novel forms of knowledge and extract neural behavior as efficient representations of knowledge to be transferred. Based on the alignment of neural behavior between both teacher and student networks during defensive distillation, the latter can achieve better learning results than other distillation-based defenses (Li et al. 2021; Xia et al. 2022).
- We optimize original training samples into pseudo samples that can induce student network to exhibit backdoor neural behavior. On this basis, the backdoor in the student network can be further removed actively when combined with a neural behavioral alignment mechanism.
- We conduct extensive experiments on a number of well-known backdoor attacks. The experimental results corroborate the effectiveness and generality of our approach.

## Related work
### Backdoor attack

We refer to a neural network that has been implanted with a backdoor as a backdoored network, and refer to a sample that has been injected with a trigger as a poisoned sample. The backdoored network exhibits backdoor behavior when it takes poisoned sample as input, namely make specific wrong prediction.

Existing backdoor attack can be divided into poison-label attack and clean-label attack according to whether the label of the poisoned sample is modified. Poison-label attack require the attacker to modify both the samples and the labels, so that the mapping between the trigger and the target label can be directly established. BadNets (Gu et al. 2017) is the first and most representative poison-label attack. The subsequent poison-label attacks are intended to improve the BadNets from the perspective of trigger design (Liu et al. 2018, 2020), trigger implanting (Chen et al. 2017), and others. The clean-label attack (Turner et al. 2019; Barni et al. 2019) is designed to solve the phenomenon of inconsistent semantics of poisoned samples and labels, and these methods often need to add additional constraints on samples from target labels.

In this paper, we choose well-known methods from clean-label attack and poison-label attack for experiment, so as to fully illustrate the generality and effectiveness of NBA.

**Backdoor defense**

Existing defense schemes can be divided into certified defenses and empirical defenses. Certified defenses (Weber et al. 2020; Jia et al. 2022) can theoretically ensure a certain degree of robustness, but their assumptions tend to be strong, they are not as effective as empirical defenses in practical situations. According to purpose and object of defense, empirical defenses can be classified into four categories, including (1) poisoned sample detection (Zeng et al. 2021; Hayase et al. 2021), (2) trigger (injected into sample) invalidation (Qiu et al. 2021; Doan et al. 2020), (3) network detection (Xu et al. 2021; Zheng et al. 2021), and (4) backdoor (implanted into network) removal (Liu et al. 2018; Wu and Wang 2021). Since the purpose of the defense is to prevent the poisoned sample from activating the backdoor, the defense only needs to be implemented on either side of the input and the model. In the first two types of methods, the input side is protected by detecting the poisoned sample or by destroying triggers in the input. The latter two types of methods defend on the model side by detecting the backdoored network or removing backdoors in it.

We argue that a backdoor attack stems from the backdoor implanted in the model, thus a defense scheme that removes the backdoor can effectively solve the problem of backdoor attacks. In general, NBA aim at eliminating backdoor from the backdoored network. Based on our proposed neural behavior alignment and pseudo-poisoned sample, NBA can further remove backdoors while improving the benign performance of the backdoored network.

**Knowledge distillation**

As a classic deep learning technique, knowledge distillation is often used in the fields of neural network quantization and transfer learning. In knowledge distillation, a well-trained network is usually used as a teacher network, and a network that lacks training is called a student network. The teacher network guides the student network to learn, and study have shown that under this learning paradigm, the student network can achieve better results than learning by itself (Hinton et al. 2015). In most scenarios where knowledge distillation is used, the structure of the teacher network will be more complex than that of the student network, but the study of Furlanello et al. (2018) shows that the student network can even achieve better performance than the teacher network when thet have the same architecture. Hinton et al. (2015) first introduced knowledge distillation in deep learning, and they used soften predictions as the knowledge to be transferred. After that, there is a lot of work to improve the efficiency of knowledge distillation by designing new knowledge to be transferred. Representative improvement works include using intermediate feature (Romero et al. 2015; Zagoruyko and Komodakis 2017), using relationship feature (Yim et al. 2017; Park et al. 2019), using structure feature (Liu et al. 2020; Xixia et al. 2020), etc.

Based on Furlanello et al. (2015), Hinton et al. (2018), we argue that knowledge distillation can be applied to backdoor removal. Existing work confirms this, and they have achieved good results in defensive distillation with attention maps (Li et al. 2021) and corresponding improvements (Xia et al. 2022). Accordingly, defensive distillation may provide a promising method of defending against backdoors. In addition, Ge et al. (2021) have considered the backdoor failure problem that may be caused by knowledge distillation, and proposed targeted optimization. However, since its threat model and attack scenarios are not consistent with those discussed in this article, we will not analyze it.

**Threat model**

We consider a common scenario, where the training process of network is outsourced to a third party. It applies to the case where the user downloads the trained model directly from the Internet or customizes the trained network through MLaaS.

The attacker is free to implant backdoors into the network in any manner he chooses. Different trigger patterns can be designed, different labels can be set, and poisoning rates can be set arbitrarily. Here, we assume that the network was successfully implanted with a backdoor and returned to the user. The user is often provided with a partially clean dataset so that they can confirm the usability of the network once it has been returned (or released). The network is expected to perform well on this dataset.

There are no details about the training process and attack methods available to the defender. He is only provided with a trained network and a small portion of clean dataset. Due to the fact that it is unknown whether a given network contains a backdoor, the proposed defense method should be network-agnostic. The defensive solution should remove the backdoor from a given network without significantly degrading the its normal performance if it is a backdoored network. Particularly if the network is clean, the defense mechanism should not significantly affect its performance.

## Methodology
### Overview

Figure 2 illustrates the pipeline of using NBA for backdoor defense. It consists of two steps: first, fine-tuning the backdoor network in order to obtain the teacher network, and then, through defensive distillation, aligning the neural behavior of the student network to remove the backdoor.

As shown in Fig. 2a, the defender fine-tunes the given network using a local clean dataset in order to obtain the teacher network. Figure 2b illustrates the subsequent defensive distillation step. As it was originally proposed for the purpose of neural network compression, knowledge distillation will not provide good results when it is applied directly to the backdoor defense. To obtain satisfactory defense performance, we optimize knowledge representation and training samples in knowledge distillation in accordance with backdoor defense features.

The improvements we have made to defensive distillation have been inspired by real-life teaching experiences. As an analogy, we compare the behavior of the backdoor network in processing samples to that of students in solving problems. Consequently, knowledge distillation can be viewed as the process by which teachers instruct students in the proper method of resolving problems. Two lessons can be drawn from practical teaching experience.

Firstly, teachers should provide students with a complete understanding of problem solving, including intermediate steps, intermediate answers, and final solutions. The student will not be able to fully comprehend the ins and outs of the correct method of solving problems if any or all of these are missing. In order to simulate this process, NBA extracts and aligns three kinds of neural behavioral within student and teacher networks. Each of these three neural behaviors within neural network has its own focus, and the combination can facilitate comprehensive learning by the student network of the knowledge imparted by the teacher network. Secondly, teachers will provide concentrated problem-solving guidance for students' error-prone problems and correct the students' faulty problem-solving methods. Inspired by this, we constructed pseudo samples to induce the student network to actively exhibit backdoor neural behaviors, aligning them with the benign neural behavior of the teacher network to remove the backdoor more effectively.

Based on this, we propose learning distillation loss and unlearning distillation loss, which are used to encourage

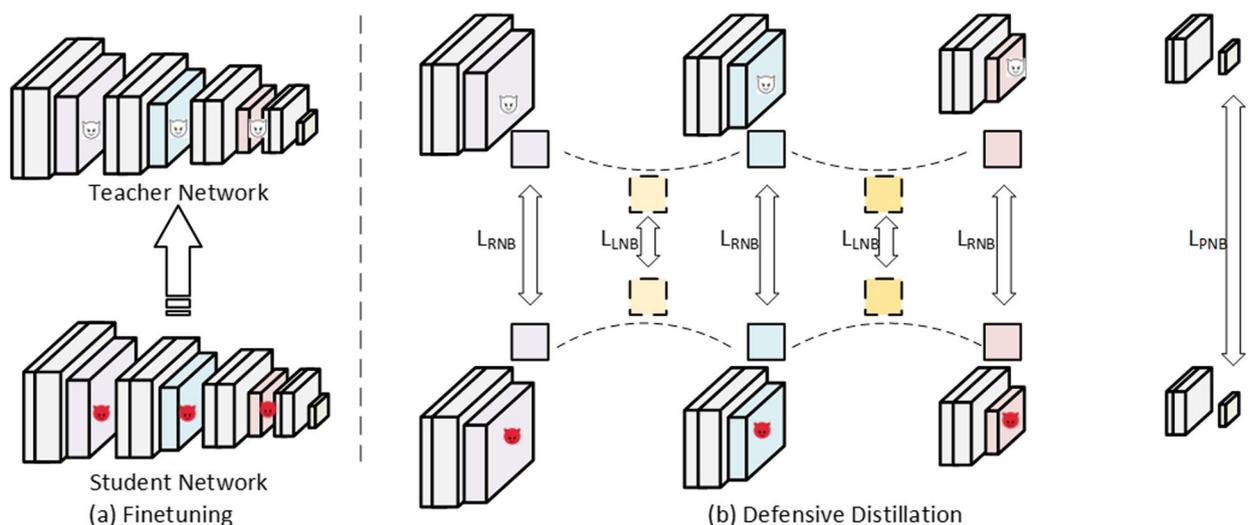

**Fig. 2** The overview of NBA. NBA consists of two main procedures to remove backdoor: (1) fine-tuning: fine-tuning based on local clean data to obtain the teacher network; (2) defensive distillation: extracting and aligning high-level representation of neural behavior from teacher network and student network. Backdoor will be eliminated from student network by optimizing the two kinds of distillation loss functions adopted in defensive distillation phase

the student network to learn benign knowledge efficiently and to actively unlearn backdoor knowledge.

### Neural behavior

Definition and extraction of the neural behavior of the neural network are the keys to NBA. We define two types of neural behaviors, namely, response neural behavior and learning neural behavior, respectively, for the intermediate answers and intermediate steps in the problem-solving process. Figure 3 shows the procedure of there two kinds of extracting neural behavior. In addition, we introduce the dark knowledge proposed by Hinton et al. (2015) as prediction neural behavior to represent the final answer of the problem.

### Response neural behavior

Previous studies (Zagoruyko and Komodakis 2017; Li et al. 2021; Xia et al. 2022; Romero et al. 2015) have shown that feature maps can represent the response of neurons inside the network to input samples. We extracted the feature maps of each intermediate layer of the network and regarded it as the original representation of the response neural behavior of the model. In order to capture the focus of the response neural behavior, inspired by Gatys et al. (2016), we use the gram matrices to capture the key features of the feature maps. In particular, the gram matrices that we get here are called response matrices, and they can be used as the high-level representation of the responsive neural behavior.

The feature maps of in the $l$-th layer of the teacher network and the student network are denoted by

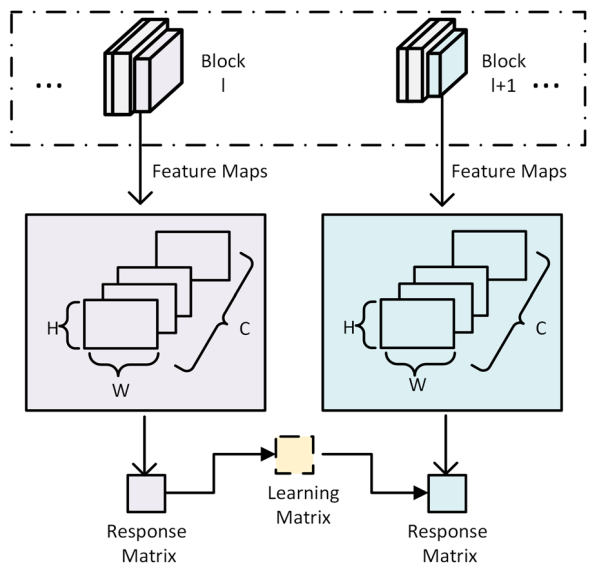

**Fig. 3** The illustration of extracting response neural behavior and learning neural behavior

$F^{Tl} \in \mathbb{R}^{C_l \times H_l \times W_l}$ and $F^{Sl} \in \mathbb{R}^{C_l \times H_l \times W_l}$, where $C_l$, $H_l$, $W_l$ represent the the number, height and width of feature maps in $l$-th layer. Further, the response matrices in the $l$-th layer of the teacher network and the student network can be denoted as $G^{Fl} \in \mathbb{R}^{c \times c}$ and $G^{Sl} \in \mathbb{R}^{c \times c}$. Response matrix $G^l$ is the inner product of the feature maps of the corresponding neural network:

$$G^l_{ij} = \sum_{i=k}^{H_l \times W_l} F^l_{ik} F^l_{jk}. \tag{1}$$

$\mathcal{L}_{RNB}$ is defined by the mean squared error between the response matrices $G^{Fl}$ and $G^{Sl}$, $l = 1, 2, \ldots, n$:

$$\mathcal{L}_{RNB} = \sum_{l=1}^{n} \mathcal{L}^l_{rnb}, \tag{2}$$

where $\mathcal{L}^l_{RNB}$ is defined as

$$\mathcal{L}^l_{RNB} = \frac{1}{4C_l^2 (H_l \times W_l)^2} \sum_{i=1}^{H_l \times W_l} \sum_{j=1}^{H_l \times W_l} \left(G^{Fl}_{ij} - G^{Sl}_{ij}\right)^2. \tag{3}$$

By optimizing the Eq. (2), student network is encouraged to align its own response neural behavior with these of the teacher network.

### Learning neural behavior

Learning neural behavior is used to simulate the intermediate problem-solving steps in actual teaching. Learning neural behavior is defined as the transformation between response neural behavior from adjacent layers, which is defined as:

$$M^l = \|G^l - G^{l+1}\|_2^2, \tag{4}$$

when $M^l$ is learning matrix between $l$-th layer and $l+1$-th layer. $M^{Tl}$ and $M^{Sl}$ can be calculated by the above equation.

Once there exists $n$ response matrices, there will be $n-1$ learning matrices. By optimizing the cross entropy loss function as follows, we encourage the student network to aligning its own learning neural behavior with teacher's:

$$\mathcal{L}_{LNB} = \sum_{l=0}^{n-1} \|M^{Tl} - M^{Sl}\|_2^2 \tag{5}$$

### Prediction neural behavior

We introduce the dark knowledge proposed by Hinton as the prediction neural behavior. Usually, the output of the model is obtained by processing the logits information through softmax. The neural behavior that can be represented in this output is limited. To this end, we follow the procedure of Hinton et al. (2015), introducing a temperature $T$ to soften the result of the softmax as the prediction neural behavior. The prediction neural behavior for class $i$ can be calculated by the follows:

$$p_i(z_i, T) = \frac{exp(z_i/T)}{\sum_{j=0}^{k} exp(z_j/T)}, \quad (6)$$

where $z_i$ and $z_j$ are logits, $k$ is the number of classes. Here we set $T = 5$.

Therefore, the alignment of prediction neural behavior can be implemented by optimizing the KL-divergence between the prediction neural behavior of the teacher network and the student network:

$$\begin{aligned}\mathcal{L}_{PNB} &= L_{KD}(p(u,T), p(z,T)) \\ &= \sum_{i=0}^{k} -p_i(u_i, T) log(p_i(z_i, T)),\end{aligned} \quad (7)$$

where $u$ and $z$ are the logits of teacher network and student network respectively.

By optimizing the $\mathcal{L}_{PNB}$, the student model is encouraged to align its own prediction neural behavior with that of the teacher.

**Learning distillation loss**
Based on the three losses defined in "Neural behavior" section, we define NBA learning distillation loss $\mathcal{L}_{LDL}$ to encourage the student network to fully learn the knowledge transferred by the teacher network during defensive distillation:

$$\begin{aligned}\mathcal{L}_{LDL} = \mathbb{E}_{(x,y)\sim\mathcal{D}}[&\lambda_1 \mathcal{L}_{RNB}(T(x), S(x)) \\ &+ \lambda_2 \mathcal{L}_{LNB}(T(x), S(x)) \\ &+ \lambda_3 \mathcal{L}_{PNB}(T(x), S(x))],\end{aligned} \quad (8)$$

where $\mathcal{D}$ is local dataset, $T$ and $S$ are teacher and student networks. Specifically, $T(\cdot)$ and $S(\cdot)$ represent different knowledge form in different loss. $\lambda_i (i = 1, 2, 3)$ are hyperparameters controlling the weights of each loss item. They are set as 2.0, 2.0 and 0.1, respectively.

**Unlearning distillation loss**
Unlearning distillation loss is proposed to correct the backdoor behavior of the student network and improve its generalization. The core idea behind it is to construct pseudo samples, and input the original samples and pseudo samples into the teacher network and student network respectively. By optimizing this loss function, the student network will further eliminate the backdoor behavior, that is, actively unlearn the backdoor knowledge.

Student network typically only show backdoor behavior in the presence of poisoned samples, but defenders only have clean samples. We introduce adversarial attacks to address this challenge.

Typically, the backdoored network can be obtained by optimizing the following loss function during training:

$$\mathcal{L} = \mathbb{E}_{(x,y)\sim\mathcal{D}_c}[l(f_\theta(x), y)] + \mathbb{E}_{(x,y)\sim\mathcal{D}_p}[l(f_\theta(x + \triangle), y_t)], \quad (9)$$

where $l(\cdot)$ denotes the loss function such as cross-entropy loss, $\mathcal{D}_c$ and $\mathcal{D}_p$ are denote the subsets of training dataset. Particularly, $\triangle$ denotes the trigger and $y_t$ denotes the target label. The function of the second item is to implant the backdoor. Optimizing this loss function actually creates a shortcut in the network for the decision-making process of recognizing the input as the target label compared to training a clean neural network. Potential adversarial attacks are thus affected.

We conduct non target attack on the student network to craft pseudo samples $x' = x + \delta$, where $\delta$ is adversarial perturbation, which can be obtained by optimizing:

$$\max_{\delta} \mathcal{L}(f_\theta, x, y), s.t. \|\delta\|_p < \epsilon \quad (10)$$

The above equation can be solved by gradient-based adversarial methods, such as Madry et al. (2018). According to different predicted labels, pseudo samples can be divided into two categories. The samples that are predicted as target labels are pseudo-poisoned samples, and they will converge to the local extremum caused by the backdoor during the optimization. This means that $\delta$ and $\triangle$ are strongly related. Figure 4 visually shows the feature maps generated by pseudo-poisoned samples and poisoned samples. They have strong similarities, which indicates that they are both able to activate the network to exhibit similar behavior. Another type of samples obtained by the above equation are called pseudo clean samples, and their predictions are inconsistent with the target label. Unlearning distillation loss is defined as follows:

$$\mathcal{L}_{UDL} = \mathbb{E}_{(x,y)\sim\mathcal{D}}[\lambda_1 \mathcal{L}_{RNB}(T(x), S(x+\delta)) \\ + \lambda_2 \mathcal{L}_{LNB}(T(x), S(x+\delta)) \\ + \lambda_3 \mathcal{L}_{PNB}(T(x), S(x+\delta))], \quad (11)$$

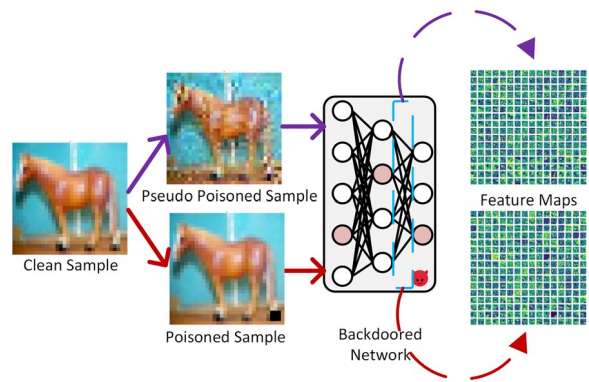

**Fig. 4** Visualization of feature maps extracted from backoored network

It is important to note that during the optimization of the above equation, the two types of pseudo samples have varying roles, but both can contribute to the enhancement of the defense performance. In particular, the pseudo-poisoned samples induce the student network to exhibit backdoor behavior, allowing backdoor knowledge to be removed more efficiently. While the pseudo-clean samples essentially provide a regularization function.

### Total loss

Overall objective of defensive distillation has the form of:

$$\mathcal{L}_{Total} = \alpha(\mathcal{L}_{LDL} + \beta \mathcal{L}_{UDL}) + l(f_\theta(x), y), \quad (12)$$

where $\alpha$ controls the weight of distillation loss in total loss, and $\beta$ controls the weight of unlearning distillation loss in distillation loss. We set $\alpha = 1.0$ and $\beta = 0.5$ in this paper.

## Experiments
### Experimental settings
#### Attack setups

We conduct experiments on 6 representative backdoor attacks, which have their own distinct characteristics in trigger design (BadNets (Gu et al. 2017), TrojanNN (Liu et al. 2018) and Refool (Liu et al. 2020)), trigger injection (Blend (Chen et al. 2017)), and label modifying (CLA (Turner et al. 2019) and SIG (Barni et al. 2019)). The poisoned samples constructed by there methods are shown in Fig. 5. We follow the settings given in the original paper, such as trigger patterns and target labels. We evaluate all attacks and defenses on CIFAR10 and GTSRB, with WideResNet (WRN-16-1) (Zagoruyko and Komodakis 2016). Other attack details are described in Table 1.

#### Defense setups

We compare our approach with the state-of-the-art defenses, including Fine-tuning (FT) (Papernot et al. 2016), Fine-pruning (FP) (Liu et al. 2018), Mode Connectivity Repair (MCR) (Zhao et al. 2020), Neural Attention Distillation (NAD) (Li et al. 2021), and Attention Relation Graph Distillation (ARGD) (Xia et al. 2022). The defenses can be compared fairly since they are based on the same threat model. Consistent with previous work (Li et al. 2021; Xia et al. 2022), we assume that the defender has a 5% clean training set. We set batch size 64, initial learning rate 0.1 and train network using the Stochastic Gradient Descent (SGD) optimizer with a momentum of 0.9.

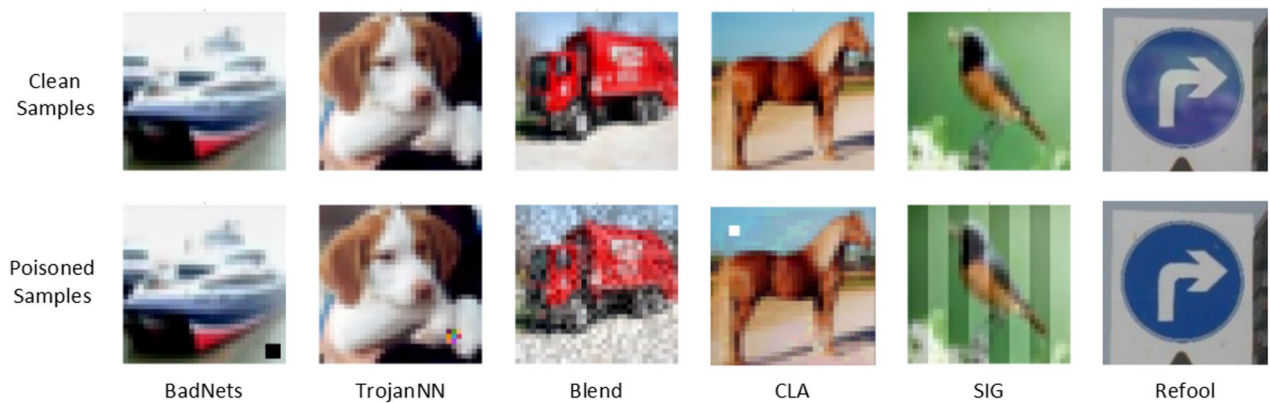

**Fig. 5** Examples of clean samples (top) and poisoned samples (bottom) used in our experiments

**Table 1** Settings of 6 well-known backdoor attacks

| Attack methods | BadNets | TrojanNN | Blend | CLA | SIG | Refool |
| --- | --- | --- | --- | --- | --- | --- |
| Dataset | CIFAR-10 | CIFAR-10 | CIFAR-10 | CIFAR-10 | CIFAR-10 | GTSRB |
| Poisoning rate | 0.1 | 0.05 | 0.1 | 0.08 | 0.08 | 0.08 |
| Trigger size | Local pattern | Local pattern | Global pattern | Local pattern | Global pattern | Global pattern |
| Target label | 0 | 0 | 0 | 0 | 0 | 0 |

*Metrics*

We use attack success rate (ASR) and benign Accuracy (BA) to evaluate the effectiveness of the defenses. In particular, the lower the ASR and the higher the BA, the better the defense method.

- *Attack Success Rate (ASR)* This metric measures the proportion of poisoned testing set predicted to the target class.
- *Benign Accuracy (BA)* This metric measures the proportion of clean testing set predicted the ground-truth classes.

**Experimental results**
*Effectiveness of NBA*

We present the detailed results on the comparison of performance in Table 2.

Overall, our proposed method achieves good defensive performance. We further illustrate this from two aspects.

We first analyze the defensive effects of NBA against different attacking methods. It is noted that NBA is always effective on different attack methods, i.e., lower ASR and higher BA can be achieved. This demonstrates NBA's impressive adaptability.

Second, we conducted a comparison between the defenses. FT and three methods based on knowledge distillation (NAD, ARGD, and NBA) achieve better results. The results of a further analysis revealed two important findings. (1) The scheme of knowledge distillation is better than FT. It is due to the fact that the latter relies solely on loss functions such as cross-entropy loss for self-learning, while the former introduces distillation loss and can benefit from the guidance of a teacher network. (2) In the internal comparison of distillation schemes, NBA achieves the best performance (average ASR is 1.52, and BA is 81.14). The reason for this is that NAD's distillation loss is only based on attention maps extracted from feature maps, and ARGD's distillation loss takes into account the order relationship between attention maps. Essentially, they are equivalent to the response neural behavior. NBA, however, adopts two different types of distillation loss simultaneously. The first type of distillation loss is learning distillation loss, which utilizes three types of neural behavior, including response neural behavior, as the form of knowledge, and is capable of leading to better learning results. There is also the unlearning distillation loss, which is capable of actively reducing backdoor neural behaviors. As a result, NBA has a significant advantage when it comes to reducing ASR and maintaining BA.

In addition, it is worth noting that although ARGD performs better than NAD on average, its BA value (80.35) is lower than NAD's (80.47) when defending against attacks such as BadNets. This indicates that the improvement achieved by ARGD is limited. In contrast, NBA outperforms both ARGD and NAD in terms of defense performance, whether defending against specific attack methods or on average. In terms of the degree of improvement, NBA is able to consistently optimize the defense performance (including BA and ASR) by at least 1 percentage point at the margin, achieving the best defense performance.

Furthermore, we find that although FT, NAD, and ARGD do not adopt a loss function similar to NBA's unlearning distillation loss, they can still reduce ASR to a certain extent. It is important, however, to stress that for these schemes, the reduction of BA relies on the "Catastrophic Forgetting" Effect (Goodfellow et al. 2014; Kirkpatrick et al. 2017) of the neural network, rather than actively removing backdoors.

**Table 2** Performance (%) comparison of 6 backdoor defenses against 6 backdoor attacks

| Attack | Defense | | | | | | | | | | | | | |
|---|---|---|---|---|---|---|---|---|---|---|---|---|---|---|
| | No defense | | Fine-tuning | | Fine-pruning | | MCR | | NAD | | ARGD | | NBA | |
| | ASR | BA | ASR | BA | ASR | BA | ASR | BA | ASR | BA | ASR | BA | ASR | BA |
| BadNets | 99.83 | 80.02 | 6.46 | 79.91 | 82.54 | 77.65 | 2.74 | 78.29 | 3.55 | 80.47 | 1.81 | 80.35 | 1.16 | 81.59 |
| TrojanNN | 99.85 | 79.95 | 5.61 | 80.03 | 52.71 | 79.96 | 25.71 | 78.68 | 3.26 | 79.58 | 2.33 | 79.97 | 1.14 | 80.42 |
| Blend | 97.83 | 82.36 | 5.23 | 79.85 | 89.12 | 80.07 | 68.85 | 79.82 | 2.77 | 81.04 | 1.16 | 81.13 | 0.85 | 80.13 |
| CLA | 98.15 | 81.14 | 7.32 | 80.06 | 35.46 | 76.88 | 17.29 | 80.03 | 8.55 | 79.64 | 5.13 | 79.96 | 1.71 | 80.24 |
| SIG | 99.62 | 82.63 | 11.29 | 80.31 | 65.31 | 80.15 | 1.80 | 79.61 | 5.69 | 80.29 | 2.14 | 80.25 | 1.95 | 81.71 |
| Refool | 96.24 | 80.37 | 8.78 | 80.24 | 59.67 | 78.22 | 8.29 | 78.25 | 4.27 | 80.01 | 4.05 | 80.04 | 2.33 | 82.72 |
| Average | 98.57 | 81.08 | 7.45 | 80.07 | 64.14 | 78.82 | 20.78 | 79.11 | 4.68 | 80.17 | 2.77 | 80.28 | 1.52 | 81.14 |

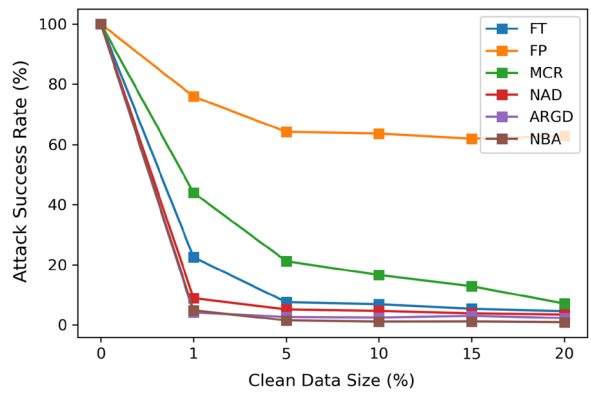

**Fig. 6** The ASRs of 6 defenses under different size of clean data

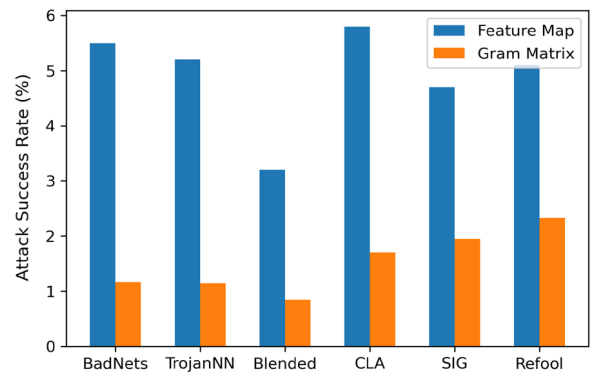

**Fig. 8** The ASRs of NBA under different representations of neural behaviors

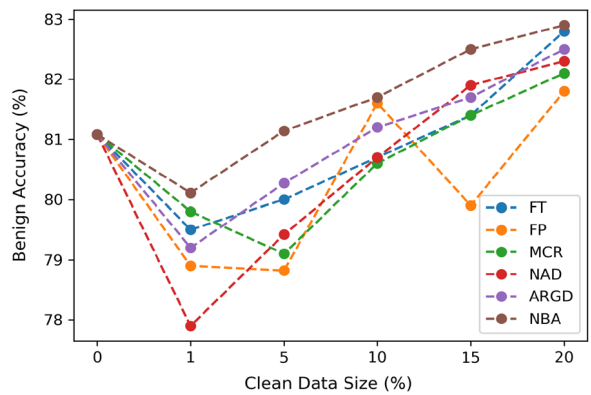

**Fig. 7** The BAs of 6 defenses under different size of clean data

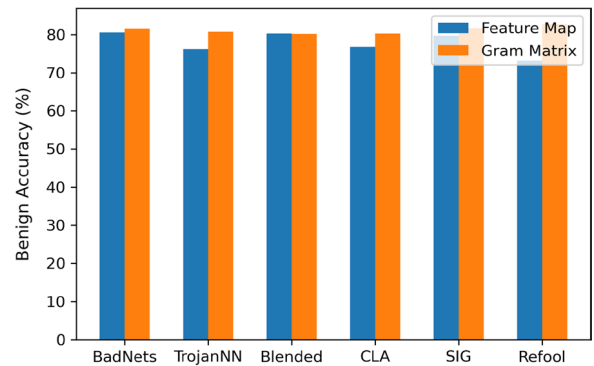

**Fig. 9** The BAs of NBA under different representations of neural behaviors

### *Effectiveness under different defender's capacity*

According to the threat model presented in this paper, defender capabilities are primarily determined by the size of the local dataset. Here we investigate the effect of local dataset size on defense performance.

Figures 6 and 7 illustrate that most defense schemes perform better as the size of the clean dataset increases. It should be noted, however, that FP is an exception. As can be seen from Fig. 6, its ASRs at 5% is similar to that at 20%. This suggests that the size of local dataset does not significantly affect the defense performance of FP. This is because more clean samples do not help FP to more accurately identify whether neurons are damaged.

With only a very small dataset (1%), NBA fails to perform as well as ARGD. However, as the dataset becomes larger, its advantages gradually become apparent. Specifically, with a dataset size of 5%, NBA achieves the best performance among all defenses.

Further observations show that the defense performance gap between these defenses except for FP narrows as the dataset size increases. This indicates that dataset size can indeed affect defense performance in a significant way. NBA, however, continues to have a significant advantage in this case. In the experimental data of 20% data set size, although NBA's BA is similar to that of other schemes, its ASR is still very low compared to other schemes.

The dataset provided by the third party with the trained networks are usually not very large at any one time. We therefore argue that our assumption of the size of the local dataset (i.e. 5% of the training set) is reasonable.

### **Further understanding of NBA**

### *The advantage of neural behavioral representations*

Here, we investigate how different neural behavior representations affect defense performance. Feature maps are treated as low-level representations in this study, while gram matrices are treated as high-level representations.

The results are shown in Figs. 8 and 9. Generally speaking, Gram matrices-based NBA has outperformed feature maps-based NBA. In spite of the fact that the feature maps represent the neural behaviors directly inside the model, there is too much redundant information within them. By contrast, Gram matrices can capture neural behavioral knowledge more effectively through inner products of feature maps, so they can be used to achieve better results of knowledge transferring.

**Table 3** Ablation results of neural behavior loss

| $\mathcal{L}_{RNB}$ | $\mathcal{L}_{LNB}$ | $\mathcal{L}_{PNB}$ | ASR | BA |
|---|---|---|---|---|
| ✓ | | | 3.41 | 83.38 |
| | ✓ | | 3.08 | 82.99 |
| | | ✓ | 3.40 | 79.37 |
| ✓ | ✓ | | 3.86 | 83.86 |
| ✓ | | ✓ | 4.32 | 82.09 |
| | ✓ | ✓ | 4.62 | 82.82 |
| ✓ | ✓ | ✓ | 2.55 | 83.91 |

**Table 4** Ablation results of distillation loss

| $\mathcal{L}_{LDL}$ | $\mathcal{L}_{UDL}$ | ASR | BA |
|---|---|---|---|
| ✓ | | 2.55 | 83.91 |
| | ✓ | 0.53 | 77.24 |
| ✓ | ✓ | 1.52 | 81.14 |

*Ablation study of distillation loss*
NBA combines three kinds of neural behavior and two forms of loss function to effectively remove backdoor. Here we perform two ablation study to demonstrate that none of them can be omitted.

In Table 3, we show that applying each neural behavior can achieve a certain defensive effect. However, when only one or two kinds of neural behaviors are used, the defenase scheme cannot achieve the best performance.

For learning distillation loss and unlearning distillation loss, we provide Table 4, which shows the performance of different settings. The results in Table 4 demonstrate that a reasonable overall defense performance can be obtained when only learning distillation loss is used. The backdoor can be removed more efficiently by unlearning distillation loss (as indicated by the lower ASR), but at the cost of the lower BA. By reasonably adjusting coefficient $\beta$ in Eq. (12), we can achieve a better trade-off between ASR and BA.

**Table 5** Performance comparison of different samples

| Settings | ASR | BA |
|---|---|---|
| NBA with only $\mathcal{L}_{RNB}$ | 2.55 | 83.91 |
| NBA with poisoned samples | 1.38 | 80.89 |
| NBA with pseudo samples | 1.52 | 81.14 |

*Possible settings for unlearning distillation loss*
Crafting pseudo samples is the key for unlearning distillation loss. The rationale and necessity of the pseudo sample crafting approach are covered in this section.

The pseudo samples are replaced with poisoned samples to perform additional experiments. Table 5 shows the experimental results. The first row presents the results of defensive distillation using only the learning distillation loss, and the corresponding ASR is reduced to 3.2. This indicates that the backdoor has been largely removed. The last two rows of the Table 5 display NBA's results using poisoned samples and pseudo samples. Both methods can further reduce the ASR (1.38 and 1.52, respectively), indicating that introducing unlearning distillation loss can indeed effectively remove the backdoor. It should be noted that there exists diminishing marginal effect in terms of removing the backdoor. There is not much difference between using poisoned samples and using pseudo samples in unlearning distillation loss. Several defense methods are capable of accurately reversing engineering the approximate poisoned sample (Qiao et al. 2019; Wang et al. 2019; Tao et al. 2022), but this is unnecessary for our approach given that even poisoned samples cannot provide a significantly better performance). In addition, these schemes require high computational overhead or attack details such as trigger size (Wang et al. 2019) when crafting potential poisoned samples.

## Conclusion

This paper presents NBA, a novel defensive distillation mechanism for backdoor removal. We optimize the knowledge distillation process from both the knowledge form and training samples to make it better suited to the defense scenario. In terms of knowledge form, we extract and align three kinds of neural behavior of networks to achieve efficient knowledge transfer. In terms of training samples, we construct pseudo samples to further eliminate backdoor from the backdoored network.

To the best of our knowledge, NBA is the first active defensive distillation mechanism and has competitive advantages in terms of backdoor removal. NBA's highly effective defense performance and realistic threat model make it an attractive candidate for practical defensive scenarios.

**Acknowledgements**
Not applicable.

**Author contributions**
The first author completed the main work of the paper and drafted the manuscript. Thesecond author participated in problem discussions and improvements of the manuscript. Both authors read and approved the final manuscript.

**Authors' Information**
Bin Wu is a faculty member at Institute of Information Engineering, Chinese Academy of Sciences. Prior to that, he obtained his Ph.D. degree from Institute of Software, Chinese Academy of Sciences 2010 under the supervision of Dengguo Feng.

**Funding**
This work was supported by the National Natural Science Foundation of China under Grant No.62272007, the National Natural Science Foundation of China under Grant No.U1936119 and the Major Science and Technology Project of Hainan Province under Grant No.ZDKJ2019003.

**Availability of data and materials**
The datasets used for the experiments are freely available to researchers. The links to the data have been cited as references.

## Declarations

**Competing interests**
The authors declare that they have no competing interests.

Received: 12 December 2022   Accepted: 29 March 2023
Published online: 03 July 2023

## References

Barni M, Kallas K, Tondi B (2019) A new backdoor attack in CNNs by training set corruption without label poisoning. In: 2019 IEEE international conference on image processing, ICIP 2019, Taipei, China, 22–25 Sep 2019. pp 101–105. IEEE

Berman DS, Buczak AL, Chavis JS, Corbett CL (2019) A survey of deep learning methods for cyber security. Information 10(4):122

Chen X, Liu C, Li B, Lu K, Song D (2017) Targeted backdoor attacks on deep learning systems using data poisoning. CoRR arXiv:1712.05526

Costa-jussà MR (2018) From feature to paradigm: deep learning in machine translation (extended abstract). In: L Jang (eds) Proceedings of the Twenty-Seventh international joint conference on artificial intelligence, IJCAI 2018, Stockholm, Sweden, 13–19 July 2018. pp 5583–5587. ijcai.org

Doan BG, Abbasnejad E, Ranasinghe DC (2020) Februus: input purification defense against trojan attacks on deep neural network systems. In: ACSAC '20: annual computer security applications conference, virtual eventl, Austin, TX, USA, 7–11 Dec, 2020, pp 897–912. ACM

Furlanello T, Zachary CL, Tschannen M, Itti L, Anandkumar A (2018) Born-again neural networks. In: Dy JG and Krause A (eds), Proceedings of the 35th international conference on machine learning, ICML 2018, Stockholmsmässan, Stockholm, Sweden, 10–15 July 2018, vol 80 of Proceedings of machine learning research, pp 1602–1611. PMLR

Gao Y, Doan BG, Zhang Z et al (2020) Backdoor attacks and countermeasures on deep learning: a comprehensive review. CoRR, arXiv: abs/2007.10760

Gatys LA, Ecker AS, Bethge M (2016) Image style transfer using convolutional neural networks. In: 2016 IEEE conference on computer vision and pattern recognition, CVPR 2016, Las Vegas, NV, USA, 27–30 June, 2016, pp 2414–2423. IEEE Computer Society

Geoffrey EH, Oriol V, Jeffrey D (2015) Distilling the knowledge in a neural network. CoRR arXiv:1503.02531

Ge Y, Wang Q, Zheng B et al (2021) Anti-distillation backdoor attacks: backdoors can really survive in knowledge distillation. In: Shen HT, Zhuang Y, Smith JR et al (eds) MM '21: ACM multimedia conference, virtual event, China, 20–24 Oct 2021, pp 826–834. ACM

Goodfellow IJ, Mirza M, Da X, Courville AC, Bengio Y (2014) An empirical investigation of catastrophic forgetting in gradient-based neural networks. In: Bengio Y and LeCun Y (eds) 2nd international conference on learning representations, ICLR 2014, Banff, AB, Canada, 14–16 April 2014, conference track proceedings

Grigorescu SM, Trasnea B, Cocias TT, Macesanu G (2020) A survey of deep learning techniques for autonomous driving. J Field Robot 37(3):362–386

Gu T, Dolan-Gavitt B, Garg S (2017) Badnets: identifying vulnerabilities in the machine learning model supply chain. CoRR arXiv: 1708.06733

Guowen Xu, Li Hongwei, Ren Hao, Yang Kan, Deng Robert H (2019) Data security issues in deep learning: attacks, countermeasures, and opportunities. IEEE Commun Mag 57(11):116–122

Hayase J, Kong W, Somani R, Oh S (2021) SPECTRE: defending against backdoor attacks using robust statistics. CoRR arXiv:2104.11315

Hu G, Yang Y, Yi D et al (2015) When face recognition meets with deep learning: an evaluation of convolutional neural networks for face recognition. In: 2015 IEEE international conference on computer vision workshop, ICCV Workshops 2015, Santiago, Chile, 7–13 Dec 2015, pp 384–392. IEEE Computer Society

Jia J, Liu Y, Cao X, Gong NZ (2022) Certified robustness of nearest neighbors against data poisoning and backdoor attacks. In: Thirty-Sixth AAAI conference on artificial intelligence, AAAI 2022, Thirty-Fourth conference on innovative applications of artificial intelligence, IAAI 2022, The Twelveth symposium on educational advances in artificial intelligence, EAAI 2022 Virtual Event, February 22–March 1, 2022, pp 9575–9583. AAAI Press, USA

Kirkpatrick James, Pascanu Razvan, Rabinowitz Neil et al (2017) Overcoming catastrophic forgetting in neural networks. Proc Natl Acad Sci 114(13):3521–3526

Koehn P (2020) Neural machine translation. Cambridge University Press, Cambridge

Li Y, Lyu X, Koren N et al (2021) Neural attention distillation: erasing backdoor triggers from deep neural networks. In: 9th international conference on learning representations, ICLR 2021, Virtual Event, Austria, 3–7 May 2021. OpenReview.net

Liu Ximeng, Xie Lehui, Wang Yaopeng et al (2021) Privacy and security issues in deep learning: a survey. IEEE Access 9:4566–4593

Liu K, Dolan-Gavitt B, Garg S (2018) Fine-pruning: defending against backdooring attacks on deep neural networks. In: Bailey M, Holz T, Stamatogiannakis M and Ioannidis S (eds) Research in attacks, intrusions, and defenses—21st international symposium, RAID 2018, Heraklion, Crete, Greece, 10–12 Sep 2018, Proceedings, vol 11050 of Lecture Notes in Computer Science, pp 273–294. Springer

Liu K, Dolan-Gavitt B, Garg S (2018) Fine-pruning: defending against backdooring attacks on deep neural networks. In: Bailey M, Holz T, Stamatogiannakis M and Ioannidis S (eds), Research in attacks, intrusions, and defenses—21st international symposium, RAID 2018, Heraklion, Crete, Greece, 10–12 Sep, 2018, Proceedings, vol 11050 of Lecture Notes in Computer Science, pp 273–294. Springer

Liu Y, Ma S, Aafer Y et al (2018) Trojaning attack on neural networks. In: 25th annual network and distributed system security symposium, NDSS 2018, San Diego, California, USA, 18–21 Feb 2018. The Internet Society

Liu Y, Ma X, Bailey J, Lu F (2020) Reflection backdoor: a natural backdoor attack on deep neural networks. In: Vedaldi A, Bischof H, Brox H and Frahm J-M (eds) Computer vision—ECCV 2020—16th European conference, Glasgow, UK, 23–28 Aug 2020, Proceedings, Part X, vol 12355 of Lecture Notes in Computer Science, pp 182–199. Springer, Berlin

Liu Y, Shu C, Wang J, Shen C (2020) Structured knowledge distillation for dense prediction. IEEE Trans Pattern Anal Mach Intell

Madry A, Makelov A, Schmidt L, Tsipras D, Vladu A (2018) Towards deep learning models resistant to adversarial attacks. In: 6th international conference on learning representations, ICLR 2018, Vancouver, BC, Canada, April 30–May 3, 2018, conference track proceedings. OpenReview.net

Muhammad K, Ullah A, Lloret J, Ser JD, de Albuquerque VHC (2021) Deep learning for safe autonomous driving: current challenges and future directions. IEEE Trans Intell Transp Syst 22(7):4316–4336

Papernot N, McDaniel PD, Wu X, Jha S, Swami A (2016) Distillation as a defense to adversarial perturbations against deep neural networks. In: IEEE symposium on security and privacy, SP 2016, San Jose, CA, USA, 22–26 May, 2016

Park W, Kim D, Lu Y, Cho M (2019) Relational knowledge distillation. In: IEEE conference on computer vision and pattern recognition, CVPR 2019, Long Beach, CA, USA, 16–20 June 2019, pp 3967–3976. Computer Vision Foundation/IEEE

Qiao X, Yang Y, Li H (2019)Defending neural backdoors via generative distribution modeling. In: Wallach HM, Larochelle H, Beygelzimer H et al (eds) Advances in neural information processing systems 32: annual conference on neural information processing systems 2019, NeurIPS 2019, 8–14 Dec 2019, Vancouver, BC, Canada, pp 14004–14013

Qiu H, Zeng Y, Guo S et al (2021) Deepsweep: an evaluation framework for mitigating DNN backdoor attacks using data augmentation. In: Cao J, Au MH, Lin Z and Yung M (eds) ASIA CCS '21: ACM Asia conference on computer and communications security, virtual event, Hong Kong, 7–11 June 2021, pp 363–377. ACM

Ribeiro M, Grolinger K, Capretz Miriam AM (2015) Mlaas: machine learning as a service. In: Li T, Kurgan LA, Palade V et al (eds), 14th IEEE international conference on machine learning and applications, ICMLA 2015, Miami, FL, USA, 9–11 Dec 2015

Romero A, Ballas N, Samira EK et al (2015) Fitnets: hints for thin deep nets. In: Bengio Y and LeCun Y (eds) 3rd international conference on learning representations, ICLR 2015, San Diego, CA, USA, 7–9 May 2015, conference track proceedings

Romero A, Ballas N, Samira EK et al (2015) Fitnets: hints for thin deep nets. In: Bengio Y and LeCun Y (eds), 3rd international conference on learning representations, ICLR 2015, San Diego, CA, USA, 7–9 May 2015, conference track proceedings

Rudin Cynthia (2019) Stop explaining black box machine learning models for high stakes decisions and use interpretable models instead. Nat Mach Intell 1(5):206–215

Tao G, Shen G, Liu Y et al (2022) Better trigger inversion optimization in backdoor scanning. In: IEEE/CVF conference on computer vision and pattern recognition, CVPR 2022, New Orleans, LA, USA, 18–24 June 2022, pp 13358–13368. IEEE

Turner A, Tsipras D, Madry A (2019) Label-consistent backdoor attacks. CoRR arXiv:1912.02771

Wang H, Guo L (2021) Research on face recognition based on deep learning. In: 3rd international conference on artificial intelligence and advanced manufacture, AIAM 2021, Manchester, UK, 23–25 Oct, 2021, pp 540–546. IEEE

Wang B, Yao Y, Shan S et al (2019) Neural cleanse: identifying and mitigating backdoor attacks in neural networks. In: 2019 IEEE symposium on security and privacy, SP 2019, San Francisco, CA, USA, 19–23 May 2019, pp 707–723. IEEE

Weber M, Xu X, Karlas B, Zhang C, Li B (2020) RAB: provable robustness against backdoor attacks. CoRR arXiv:2003.08904

Wu D, Wang Y (2021) Adversarial neuron pruning purifies backdoored deep models. In: Ranzato M, Beygelzimer A, Dauphin YN, Liang P and Vaughan JW (eds), Advances in Neural Information Processing Systems 34: annual conference on neural information processing systems 2021, NeurIPS 2021, 6–14 Dec 2021, virtual, pp 16913–16925

Xia J, Wang T, Ding J, Wei X, Chen M (2022) Eliminating backdoor triggers for deep neural networks using attention relation graph distillation. In: De Raedt L (eds), Proceedings of the Thirty-First international joint conference on artificial intelligence, IJCAI 2022, Vienna, Austria, 23–29 July 2022, pp 1481–1487. ijcai.org

Xixia Xu, Zou Qi, Lin Xue, Huang Yaping, Tian Yi (2020) Integral knowledge distillation for multi-person pose estimation. IEEE Signal Process Lett 27:436–440

Xu X, Wang Q, Li H et al (2021) Detecting AI trojans using meta neural analysis. In: 42nd IEEE symposium on security and privacy, SP 2021, San Francisco, CA, USA, 24–27 May 2021, pp 103–120. IEEE

Yim J, Joo D, Bae J-H, Kim J (2017) A gift from knowledge distillation: fast optimization, network minimization and transfer learning. In: 2017 IEEE Conference on computer vision and pattern recognition, CVPR 2017, Honolulu, HI, USA, 21–26 July 2017, pp 7130–7138. IEEE Computer Society

Zagoruyko S, Komodakis N (2016) Wide residual networks. In: Wilson RC, Hancock ER and Smith WAP (eds) Proceedings of the British machine vision conference 2016, BMVC 2016, York, UK, 19–22 Sep 2016. BMVA Press, UK

Zagoruyko S, Komodakis N (2017) Paying more attention to attention: improving the performance of convolutional neural networks via attention transfer. In: 5th international conference on learning representations, ICLR 2017, Toulon, France, 24–26 April 2017, conference track proceedings. OpenReview.net

Zeng Y, Park W, Mao ZM, Jia R (2021) Rethinking the backdoor attacks' triggers: a frequency perspective. In: 2021 IEEE/CVF international conference on computer vision, ICCV 2021, Montreal, QC, Canada, 10–17 Oct 2021, pp 16453–16461. IEEE

Zhao P, Chen P-Y, Das P, Ramamurthy KN, Lin X (2020) Bridging mode connectivity in loss landscapes and adversarial robustness. In: 8th International Conference on Learning Representations, ICLR 2020, Addis Ababa, Ethiopia, 26–30 April 2020. OpenReview.net

Zheng S, Zhang Y, Wagner H, Goswami M, Chen C (2021) Topological detection of trojaned neural networks. In: Ranzato MA, Beygelzimer A, Dauphin YN, Liang P and Vaughanc JW (eds), Advances in neural information processing systems 34: annual conference on neural information processing systems 2021, NeurIPS 2021, 6–14 Dec 2021, virtual, pp 17258–17272

**Publisher's Note**

Springer Nature remains neutral with regard to jurisdictional claims in published maps and institutional affiliations.